\documentclass{article}
\usepackage{newpasp}
\usepackage{epsfig}
\begin{document}
\title{The VSOP 5~GHz Continuum Survey, recent results}

\author{R. Dodson,
        H. Hirabayashi,
        P. G. Edwards,
        K. Wiik}  \affil{Institute of Space and Astronautical Science, 
                 JAXA, 3-1-1 Yoshinodai,
                 Sagamihara, Kanagawa 229-8510, Japan}

\author{E. B. Fomalont}\affil{National Radio Astronomy Observatory, 520 Edgemont Road,
                 Charlottesville, VA 22903, USA}

\author{J. E. J. Lovell}\affil{Australia Telescope National Facility,
                 Commonwealth Scientific and Industrial Research Organization,
                 P. O. Box 76, Epping NSW 2122, Australia}
\author{G. A. Moellenbrock}\affil{National Radio Astronomy Observatory, 
                 P.O. Box 0, Socorro, NM 87801, USA}
\author{W. K. Scott}\affil{Physics and Astronomy Department, University of Calgary,
                 2500 University Dr. NW,
                 Calgary, Alberta, Canada, T2N 1N4}

\begin{abstract}
In February 1997 the Japanese radio astronomy satellite Halca was
launched to provide the space-bourne element for the VSOP
mission. Approximately twenty-five percent of the mission time has
been dedicated to the VSOP Survey, a 5~GHz survey of bright, compact AGN.
We present the results from the ongoing analysis.
Both the final, calibrated, high resolution images and plots of
visibility amplitude versus {\em uv} distance for the first 102
of the sources have been prepared and has been submitted. Papers
on the methods and the models from fitting the cumulative {\em uv}
amplitudes will also be submitted. The analysis of the second half is
well underway.
\end{abstract}

\section{Introduction}

The radio astronomy satellite HALCA was launched by the former
Institute of Space and Astronautical Science (now part of Japanese
Aerospace eXpolaration Agency (JAXA))in February 1997 to participate
in Very Long Baseline Interferometry (VLBI) observations with arrays
of ground radio telescopes. It was was placed in an orbit with an
apogee height above the Earth's surface of 21,400\,km, a perigee
height of 560\,km, and an orbital period of 6.3~hours. HALCA provides
the longest baselines of the VLBI Space Observatory Programme (VSOP),
an international endeavor that has involved over 25 ground radio
telescopes, five tracking stations and three correlators (Hirabayashi
et al. 1998; 2000a).

HALCA has now passed the end of the Guaranteed Observing Time period,
and with the completion of the Memorandum of Understanding in February
2002 the NASA tracking and orbital calculation support ceased and the
observation program has turned to completing the Survey. The orbital
determination and spacecraft tracking are now completely indigenous to
Japan and ISAS.

\section{The Observations}
VSOP Survey observations use $\sim$3 ground telescopes and HALCA,
co-observing for up to $\sim$6\,hours. Ground radio telescopes that
have made significant contributions to Survey Program observations
include Ceduna (Australia), the (no longer operational) Green Bank
43\,m (USA), Hartebeesthoek (South Africa), Hobart (Australia),
Kalyazin (Russia), Kashima (Japan), Mopra (Australia), Noto (Italy),
Shanghai (China), Torun (Poland), and Usuda (Japan). As in all VSOP
observations, two 16\,MHz bandwidths of two-bit sampled, left-circular
polarization data are recorded (Hirabayashi et al.\ 2000).  Data are
usually correlated at either the Penticton correlator or the Mitaka
correlator. After correlation, the data are sent to ISAS for
distribution to the Survey Reduction Team members.  The reduction of
Survey observations is described in Moellenbrock et al.\ (2000) and
Hirabayashi et al.\ (2000).

As of January 2004, over 230 of these sources had been observed.
At the currently sustainable rate (since October 1999, when
one of HALCA's four reaction wheels stopped working) of $\sim$2
observations per week, the remaining Survey observations are expected
to be completed in mid 2004.

\section{Data Reduction}

Analysis of the data has been well described elsewhere (e.g. Lovell et
al 2004) and hence will only be briefly outlined here. The data is
imported into AIPS, amplitude calibrated (with the system temperature
and, if needed, autocorrelation normalised) then fringe fitted. After
satisfactory delay and rate calibration it was summed to a single
channel and exported to DIFMAP for model fitting and self calibration.

\section{Sample Results}
\subsection{The Cumulative Visibility Amplitudes}

Horiuchi et al (2004) model fitted the cumulative observed visibilities
to explore the structure of a 'typical' AGN. This was found to have a
resolving component, a jet and a residual non-resolved core
component. See Figure 1.

\begin{figure}
\begin{center}
\epsfig{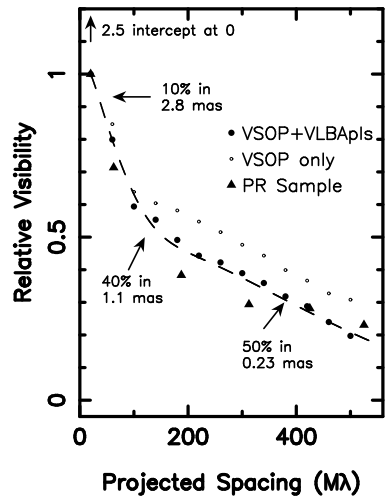}
\caption{The model fit to the cumulative visibilities, showing a
  'resolving component', a 'jet' and a non-resolved core}
\end{center}
\end{figure}

\subsection{Images of all the sources}

Scott et al (2004) covered the data reduction carried out for over 102
sources. The first detailed paper of results (``The VSOP 5GHz AGN
survey: III imaging results the first 102 sources'') has been
submitted. A typical survey source is J1837-71, a very
recently discovered GPS source (Edwards \& Tingay 2004) that had not
been imaged previously. It was observed with HALCA, HartRAO, Hobart \&
Mopra. See Figure 2.

\begin{figure}
\begin{center}
\epsfig{file=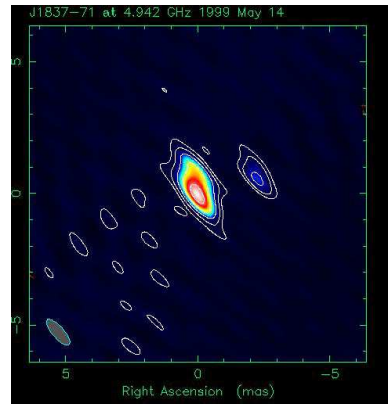,width=5cm}
\caption{An example image: the recently discovered GPS source 1837-71,
  observed as part of the VSOP survey.}
\end{center}
\end{figure}

\section{Conclusions}

The satellite continues to make survey observations, and will do so
while the satellite is functioning. These observations are being
analysed cumulatively and individually and providing interesting
results.

These results are not only important for the understanding the target
sources, mainly AGN's, but also for the planning of future space-VLBI
missions such as the VSOP-2 mission, which will have a resolution of
nearly a magnitude better. 

\section*{References}

Edwards, P., Tingay, S., 2004, A\&A, submitted\\
Hirabayashi, H., et al., 2000, PASJ, 52, 997-1014\\
Hirabayashi H. et al., 1998, Sci, 281, 1825 \\
Horiuchi et al, 2004, ApJ, submitted\\
Lovell et al, 2004, ApJ, submitted\\
Moellenbrock G. et al., 2000, APRS conf, 177 \\
Scott et al, 2004, ApJS, submitted\\

\end{document}